\documentclass[a4paper,10pt,twoside]{cpc-hepnp}
\usepackage{CJK,upgreek,fancyhdr}
\usepackage{multicol}
\usepackage{graphicx}
\usepackage{booktabs}
\usepackage{amssymb,bm,mathrsfs,bbm,amscd}
\usepackage[tbtags]{amsmath}
\usepackage{lastpage}
\usepackage{longtable}
\usepackage{changes}
\usepackage{lineno,hyperref}
\usepackage{threeparttable} 
\usepackage{multirow}
\usepackage{subfigure} 
\usepackage{ulem}
\usepackage{color}
\usepackage{amsmath}
\begin{document}
\begin{CJK*}{GB}{gbsn}

\fancyhead[c]{\small Chinese Physics C~~~Vol. xx, No. x (2022) xxxxxx}
\fancyfoot[C]{\small 010201-\thepage}


\title{Systematic calculations of cluster radioactivity half-lives in trans-lead nuclei
\thanks{This work is supported in part by the National Natural Science Foundation of China (Grants No.12175100 and No.11975132 ), the Construct Program of the Key Discipline in Hunan Province, the Research Foundation of Education Bureau of Hunan Province, China (Grants No.18A237), the Natural Science Foundation of Hunan Province, China (Grants No. 2018JJ3324), the Innovation Group of Nuclear and Particle Physics in USC, the Shandong Province Natural Science Foundation, China (Grant No. ZR2015AQ007), the National Innovation Training Foundation of China (Grant No.201910555161), and the Opening Project of Cooperative Innovation Center for Nuclear Fuel Cycle Technology and Equipment, University of South China (Grant No. 2019KFZ10).}}
\author{%
 Lin-Jing Qi (ØÁÁÖ¾²)$^{1}$ 
\quad  Dong-Meng Zhang (ÕŶ¬ÃÈ) $^{1}$
\quad  Song Luo(ÂæËÉ) $^{1}$ 
\quad Xiao-Hua Li (ÀîС»ª) $^{1,4,5;1)}$\email{lixiaohuaphysics@126.com}\\
\quad Xi-Jun Wu(Îâϲ¾ü)$^{2;2)}$\email{wuxijun1980@yahoo.cn}
\quad Chun-Tian Liang (Áº´ºÌñ)$^{3;3)}$\email{chuntianliang@hotmail.com}
}
\maketitle 
\address{%
$^1$ School of Nuclear Science and Technology, University of South China, Hengyang 421001, China\\
$^2$School of Math and Physics, University of South China, Hengyang 421001, China \\
$^3$School of Science Tianjin Chengjian University, Tianjin 300384, China\\
$^4$ Cooperative Innovation Center for Nuclear Fuel Cycle Technology $\&$ Equipment, University of South China, Hengyang 421001, China\\
$^5$ Key Laboratory of Low Dimensional Quantum Structures and Quantum Control, Hunan Normal University, Changsha 410081, China\\

}

\begin{abstract}
In the present work, based on Wentzel-Kramers-Brillouin (WKB) theory, considering the cluster preformation probability ($P_{c}$), we systematically investigate the cluster radioactivity half-lives of 22 trans-lead nuclei ranging from $^{221}$Fr to $^{242}$Cm. As for $P_{c}$, when the mass number of the emitted cluster $A_{c}$ $<$ 28, it is obtained by the exponential relationship of $P_{c}$ to the $\alpha$ decay preformation probability ($P_{\alpha}$)  proposed by R.Blendowskeis $et$ $al.$ [\href{https://doi.org/10.1103/PhysRevLett.61.1930}{Phys. Rev. Lett. \textbf{61}, 1930 (1988)}], while $P_{\alpha}$ is calculated through cluster-formation model (CFM). Whereas $A_{c}$ $\ge$ 28, it is achieved through the charge-number dependence of $P_{c}$ on the decay products proposed by Ren $et$ $al.$ [\href{https://doi.org/10.1103/PhysRevC.70.034304}{Phys. Rev. C  \textbf{70}, 034304 (2004)}]. The half-lives of cluster radioactivity have been calculated by the density-dependent cluster model [\href{https://doi.org/10.1103/PhysRevC.70.034304}{Phys. Rev. C  \textbf{70}, 034304 (2004)}] and by the unified formula of half-lives for alpha decay and cluster radioactivity [\href{https://doi.org/10.1103/PhysRevC.78.044310}{Phys. Rev. C \textbf{78}, 044310 (2008)}]. For comparison, a universal decay law (UDL) proposed by Qi $et$ $al.$ [\href{https://doi.org/10.1103/PhysRevC.80.044326}{Phys. Rev. C  \textbf{80}, 044326 (2009)}], a semi-empirical model for both $\alpha$ decay and cluster radioactivity proposed by Santhosh [\href{https://doi.org/10.1088/0954-3899/35/8/085102}{J. Phys. G: Nucl. Part. Phys. \textbf{35}, 085102 (2008)}] and a unified formula of half-lives for alpha decay and cluster radioactivity [\href{https://doi.org/10.1103/PhysRevC.78.044310}{Phys. Rev. C \textbf{78}, 044310 (2008)}] are also used.  The calculated results in our work, Ni's formula as well as UDL can well reproduce  the experimental data and are better than those in Santhosh's model. In addition, we extend this model to predict the half-lives for 51 nuclei, whose cluster radioactivity is energetically allowed or observed but yet not quantified in NUBASE2020. 
\end{abstract}



\footnotetext[0]{\hspace*{-3mm}\raisebox{0.3ex}{$\scriptstyle\copyright$}2022
Chinese Physical Society and the Institute of High Energy Physics
of the Chinese Academy of Sciences and the Institute
of Modern Physics of the Chinese Academy of Sciences and IOP Publishing Ltd}%

\begin{multicols}{2}

\section{Introduction}
\label{section 1}
In nature, radioactive nuclei translate their unstable states to stable states, based on the minimum energy principle, either by $\alpha$, $\beta$ and/or $\gamma$ emissions, or emitting particles heavier than $\alpha$ particles \cite{1,2,3} but lighter than the lightest fission fragments, generally known as cluster radioactivity \cite{4,5,6,7,56,69,68}. This subtle process, intermediate between $\alpha$ decay and spontaneous fission, undoubtedly involves vital nuclear structure information such as ground state half-life time, nuclear spin and parity, deformations of nuclear structure, shell effects and so on \cite{8,100,101}.
 In 1980, S$\check{a}$ndulescu, Poenaru and Greiner firstly predicted this type of decay mode \cite{9,10,11,12}. Before long, experiments conducted by Rose and Jones in 1984 \cite{10,12}, for observing $^{14}$C emitted particle from $^{223}$Ra, verified the realistic existence of this novel radioactivity. Soon afterwards, multiple extra elementary clusters such as $^{20}$O, $^{24}$Ne, $^{23}$F, $^{28}$Mg and $^{34}$Si have been discovered experimentally in trans-lead region \cite{14,15}, leading to the doubly magic daughter nucleus $^{208}$Pb or its neighboring nuclei. 
\par Up to now, abundant theoretical models have been proposed to deal with this radioactivity process. In general, these models can be  divided into two kinds of categories: $\alpha$-like models and fission-like models \cite{16,17,18,19}, by virtue of the intermediate characteristics. The former, just like the tunnelling theory of $\alpha$ decay \cite{60,61,62,21}, considering this process as non-adiabatic process, supposes the cluster is preformed in the parent nucleus with a certain cluster formation probability, determined by the overlapping region of both the parent nucleus and the daughter nucleus before they could penetrate the barrier with an available decay energy of cluster radioactivity $Q_{c}$ \cite{20}. For actually, in 2001, considering nuclear proximity energy and quasimolecular shape, Royer generalized the conventional liquid drop model (GLDM) \cite{19}  to systematically calculate the cluster  radioactivity half-lives. Soon later, considering the cluster preformation factor $P_{c}$ as the charge-number dependence form, Ren $et$ $al.$ employed microscopic density-dependent cluster model (DDCM) where the realistic M3Y nucleon-nucleon interaction is used to investigate this exotic decay mode \cite{9}. In 2009, extending quasi-bound wave function, firstly used in $\alpha$ decay, to cluster radioactivity, based on DDCM, Ni $et$ $al.$ systematically studied the half-lives of cluster radioactivity \cite{4}. In the later, the cluster is assumed to be formed along with the emitted process which is described as adiabatic with geometric shape constant variations from the parent nucleus during its penetration through the nuclear barrier. Based on this assumption, in 2013, Santhosh $et$ $al.$ used Coulomb and proximity potential model (CPPM) to calculate the half-lives of $\alpha$ decay and cluster radioactivity of $^{248-254}$Cf isotopes for the purpose of exploring the stability of these nuclei against these decay modes \cite{17}. Recently, Ajeet Singh $et$ $al.$ have used the effective liquid drop model (ELDM) as well as the mass excess data calculated by the relativistic mean-field (RMF) to study the cluster radioactivity half-lives \cite{66}. In addition, many phenomenological semi-empirical formulas have been proposed to deal with this phenomenon. For instance, Ni $et$ $al.$ proposed a unified formula of half-lives for both $\alpha$ decay and cluster radioactivity in 2008 \cite{22}. All of them can explicitly elaborate this exotic decay mode and provide reliable theoretical foundation for the future research.

 \par No matter in which theory, $P_{c}$ plays an indispensable role in calculating cluster radioactivity half-lives. Different models have different methods to deal with  $P_{c}$. In preformed cluster model (PCM) \cite{23,24} proposed by Gupta and Malik, $P_{c}$ is calculated through solving the stationary Schr$\ddot{o}$dinger equation for the dynamical flow of mass and charge. In terms of fission-like models, $P_{c}$ is regarded as the penetrability on the inner part of the barrier for the overlapping region. As for the unified fission model (UFM) \cite{25},  $P_{c}$ is usually considered as unity. Based on the fact that the   $\alpha$ particle is N$=$Z system with a larger binding energy, it is reasonable to deem the preformation probability just as 1 within Gamow's theory for $\alpha$ decay. Nevertheless, as in the case of cluster radioactivity, the emitted particles in the whole cluster family are completely N$\neq$Z systems. The more nucleons aggregate as a cluster, the less for the possibility exists in the parent nucleus \cite{34}. Therefore, when UFM is applied to calculate cluster radioactivity half-lives, the preformation probability, assumed to be 1 may be reevaluated. To this end,  based on  Wentzel-Kramers-Brillouin (WKB) theory, considering cluster radioactivity preformation probability $P_{c}$, we systematically investigate the cluster radioactivity half-lives of 22 trans-lead nuclei ranging from $^{221}$Fr to $^{242}$Cm, while the interaction potential between the emitted cluster and daughter nucleus in the overlapping region is a sum of the repulsive Coulomb potential $V_{C}(r)$, a modified Woods-Saxon type nuclear potential  $V_{N}(r)$  and the centrifugal potential $V_{l}(r)$. As for $P_{c}$, when $A_{c}$$<$28 we calculate it within cluster-formation model (CFM) \cite{26,41,55} combined with the famous exponential relationship of $P_{c}$ to $P_{\alpha }$ proposed by R. Blendowske and H. Walliser \cite{27}. However, with $A_{c}$ going beyond the limit this relationship may not work \cite{29}. As is clearly indicated in Fig.2 from Ref.\cite{29}, $P_{c}$ in logarithmical form keep a good linear relationship with the mass number of the emitted cluster when $A_{c}$$<$28. The curve is bent obviously when 28$<$$A_{c}\le$40 and the slope of the curve begin to decrease with the inceasing of emitted cluster mass number. Therefore, in this work, when $A_{c}\geq{28}$, $P_{c}$ is obtained by the charge-number dependence of it on the decay products in DDCM proposed by Ren $et$ $al.$ \cite{9}.

 This article is organized as follows. A brief introduction of the theoretical framework for cluster radioactivity half-life, CFM and  semi-empirical formulas is presented in Section \ref{section 2}.  Detailed numerical results and discussion are given in Section \ref{section 3}. Section \ref{section 4} is a simple summary. 
\section{Theoretical framework}
\label{section 2}
\subsection{The half-lives of the cluster radioactivity}

The half-life for the cluster radioactivity is defined as \cite{32}
\begin{equation}
T_{1/2}=\frac{\ln 2}{\lambda},             
\end{equation}
where $\lambda$ is denoted as the decay constant determined as the product of the penetration probability $P$, the assault frequency $\nu$ and the cluster-preformation probability $P_{c}$. It can be expressed as \cite{33}

\begin{equation}
\lambda=\nu PP_{c}.             
\end{equation}
Here, $P$, the penetrability of the cluster acrossing the barrier, is calculated by the Wentzel-Kramers-Brillouin (WKB) approximation.  It can be expressed as
\begin{equation}\label{3}
P=\rm{exp}{\lbrace -\frac{2}{\hbar}\int_{R_{in}}^{R_{out}}\sqrt {2\mu (V(r)-Q_{c}) }\mathrm{d}r\rbrace},
\end{equation}
where $\hbar$ is the reduced Planck constant. $\mu$ =$\frac{M_{c} M_{d}}{M_{c}+M_{d}}$ is the reduced mass of emitted cluster-daughter nucleus system with $M_{c}$ and $M_{d}$ being the masses of emitted cluster and daughter nucleus, respectively \cite{38}. $R_{in}$=$C_{1}$+$C_{2}$ \cite{34} is the saddle point for the touching configuration with
$C_{i}$=$R_{i}$(1-$\frac{b^2}{R_{i}^2}$) ($i$=1,2) being
  the S$\ddot{u}$ssmann Central radii \cite{35} of daughter and cluster nucleus on account of the surface correction to the sharp radius $R_{i}$. $b$=1 $fm$ is the diffuseness parameter of the nuclear surface taken from Ref.\cite{54}. As for sharp radius $R_{i}$, it is given by \cite{58}
\begin{equation} 
R_{i}=1.28A_{i}^{1/3}-0.76+0.8A_{i}^{-1/3},
\end{equation}
where $A_{i}$ ($i=c,d$) are the mass number of the emitted cluster and daughter nucleus, respectively.
$R_{out}$ is the outer turning point of the barrier satisfied the condition $V(R_{out})=Q_{c}$ \cite{5}. In this work, $Q_{c}$ can be obtained by 
	\begin{equation}
	Q_{c}=B(A_{c},Z_{c})+B(A_{d},Z_{d})-B(A,Z),    
	\end{equation}
	where $B(A_{c},Z_{c})$, $B(A_{d},Z_{d})$ and $B(A,Z) $ are, respectively, the binding energy of the emitted cluster, daughter and parent nuclei taken from AME2020 \cite{46} and NUBASE2020 \cite{59} with $Z_{c}$, $Z_{d}$ and $Z$ being the proton numbers of the emitted cluster, daughter and parent nucleus and $A$ being the mass number of the parent nucleus. 

 The $V(r)$ in Eq. \ref{3} is the total interacting potential between the emitted cluster and the daughter nucleus including nuclear, Coulomb and centrifugal potential barriers. It can be written as
\begin{equation}
V(r)=V_{N}(r)+V_{C}(r)+V_{l}(r),
\end{equation}
where $V_{N}$(r) is the nuclear potential. In this work, we choose it as a Woods-Saxon form \cite{18}, which can be expressed as 
\begin{equation}
V_{N}(r)=\frac{V_{0}}{1+\rm{exp}[\frac{r-R_{0}}{a}]}
\end{equation}
with 
\begin{equation}
R_{0}=r_{c}+r_{d}-1.37.
\end{equation}
 Here $r_{i} (i=c,d)$ are the nuclear charge radii. They can be expressed as
\begin{equation}
r_{i}=1.27A_{i}^{1/3},  i=c,d.
\end{equation}
The potential depth $V_{0}$ and diffuseness $a$ are parameterized as \cite{18}
\begin{equation}
V_{0}=-44.16[1-0.4(I_{d}+I_{c})]\frac{A_{d}^{1/3}A_{c}^{1/3}}{A_{d}^{1/3}+A_{c}^{1/3}},
\end{equation}
\begin{equation}
a=0.5+0.33I_{d}.
\end{equation}
Here $I_{i}$=$\frac{N_{i}-Z_{i}}{A_{i}}$ ($i=c,d$) are the relative neutron excess of the emitted cluster and daughter nucleus with $N_{i}$ and Z$_{i}$ ($i=c,d$) being the neutron numbers of the emitted cluster and daughter nucleus, respectively.
$V_{C}$ is the Coulomb potential of a uniformly charged
sphere, which can be given by 
\begin{equation}
V_{C}(r)=\frac{e^{2}Z_{c}Z_{d}}{r},
\end{equation}
where $e^2=1.4399652 MeV\cdot fm$ is the square of the  electronic elementary charge \cite{12}.

 $V_{l}$ in Eq. \ref{3} is the centrifugal potential. In this work, we 
 choose it as the Langer modified form since $l(l+1)$ $\to$ $(l+\frac{1}{2})^2$ is a necessary correction for one-dimensional problems \cite{33}. It can be written as
\begin{equation}
V_{l}(r)=\frac{\hbar^{2}(l+\frac{1}{2})^{2}}{2\mu r^{2}},
\end{equation}
where $l$ is the angular momentum carried by the emitted cluster. 
It can be obtained by

\begin{eqnarray}\label{14}
l=\left\{\begin{array}{llll}
\Delta_{j}, &\rm{for \ even \ \Delta_{j} \ and \ \pi_{p}=\pi_{d}}, \\
\Delta_{j}+1, &\rm{for \ even \ \Delta_{j} \ and \ \pi_{p}\neq\pi_{d}},\\
\Delta_{j}, &\rm{for \ odd\ \Delta_{j} \ and \ \pi_{p}\neq\pi_{d}}, \\
\Delta_{j}+1, &\rm{for \ odd \ \Delta_{j} \ and \ \pi_{p}=\pi_{d} },
\end{array}\right.
\end{eqnarray}
 where $\Delta_{j}=\lvert j_{p}-j_{d}-j_{c} \rvert$,  $j_{c},\pi_{c},j_{p},\pi_{p},j_{d},\pi_{d}$ represent the isospin and parity values of the emitted cluster, parent and daughter nuclei, respectively.

  The cluster moves back and forth inside the parent nucleus with a certain speed before penetrating the barrier. For the purpose of evaluating assaults per time unit for the ground state, $\nu$ is presented as \cite{39,57}

\begin{equation}
\nu=\frac{\pi \hbar}{2\mu R_{in}^2}.             
\end{equation}

For the preformation probability $P_{c}$,  when the mass number of the cluster $A_{c}$ $<$ 28, it can be expressed as \cite{27}
\begin{equation}
P_{c}=[P_{\alpha}]^\frac{(A_{c}-1)}{3},
\end{equation}
where $P_{\alpha}$ is the $\alpha$ decay preformation probability.
 In this work, it is obtained by CFM \cite{42}. For completeness, the detailed information about CFM will be presented in next subsection. Whereas the mass number of the cluster $A_{c}$ $\geq$ 28, the exponential relationship of $P_{c}$ to $P_{\alpha}$ may not work \cite{29}. Therefore, in this work,  $P_{c}$ is calculated through an empirical formula proposed by Ren $et.al$ \cite{9} for $A_{c}$ $\geq$ 28. It can be expressed as
\begin{eqnarray}\label{eq 17}
log_{10} P_{c}=\left\{\begin{array}{llll}
 -(0.01674Z_{c}Z_{d}-2.035466),\\
\rm{for \ even-even \ nuclei}\\
 -(0.01674Z_{c}Z_{d}-2.035466)-1.175,\\
\rm{for \ odd-A \ nuclei}.
\end{array}\right.
\end{eqnarray}

\subsection{Cluster-formation model}

  In CFM, the total initial clusterization state $\psi$ of the considered emitted cluster-daughter nucleus system is a linear combination of all its $n$ possible clusterization $\psi_{i}$
 states \cite{42}
\begin{equation}
\psi=\sum_{i}^{N}a_{i}\psi_{i},
\end{equation}
\begin{equation}
a_{i}=\int{\psi_{i}}^{*}\psi\mathrm{d}\tau,
\end{equation}
where $a_{i}$ is the superposition coefficient of $\psi_{i}$.
On account of orthogonality condition \cite{43}
\begin{equation}
\sum_{i}^{N}\lvert a_{i} \rvert^{2}=1.
\end{equation}
The total Hameiltonian $H$ is consisted of the accordingly different clusterization configuration \cite{44}. It can be expressed as
\begin{equation}
	H=\sum_{i}^{N}H_{i},
\end{equation} 
where $H_{i}$ is the $i$-th Hameiltonian of clusterization state $\psi_{i}$. On account of all the clusterizations describing the same nucleus, they are assumed as sharing the same total energy
$E$ of the total wave function \cite{43}.
Furthermore, considering the orthonormality of the clusterization wave functions, $E$ can be written as 
\begin{equation}
E=\sum_{i}^{N}\lvert a_{i} \rvert^{2}E=\sum_{i}^{N}E_{f_{i}},
\end{equation}
 where $E_{f_{i}}$ is the formation energy for the cluster in clusterization state $\psi_{i}$. Therefore,  the $\alpha$ decay preformation probability $P_{\alpha}$ can be obtained by \cite{42,44,51}: 
 \begin{equation}
 P_{\alpha}=\lvert a_{\alpha} \rvert^{2}=\frac{E_{f_{\alpha}}}{E}.
 \end{equation}
 Here $a_{\alpha}$ and $E_{f_{\alpha}}$ are, respectively, the coefficient of the $\alpha$ clusterization state and the formation energy of the $\alpha$ particle. 
The $\alpha$ formation energy $E_{f_{\alpha}}$ and total system energy $E$ can be classified as 4 different cases in the following expressions \cite{26,43}
 
 case
 $\uppercase\expandafter{\romannumeral1}$ for even-even nuclei:
 \begin{equation}
 \begin{split}
E_{f_{\alpha}}=3B(A,Z)+B(A-4,Z-2)\\
-2B(A-1,Z-1)-2B(A-1,Z),
\end{split}
 \end{equation}
\begin{equation}
E=B(A,Z)-B(A-4,Z-2);
\end{equation}

case $\uppercase\expandafter{\romannumeral2}$ for even-odd nuclei:
\begin{equation}
\begin{split}
E_{f_{\alpha}}=3B(A-1,Z)+B(A-5,Z-2)\\
-2B(A-2,Z-1)-2B(A-2,Z),
\end{split}
\end{equation}
  \begin{equation}
E=B(A,Z)-B(A-5,Z-2);
\end{equation}

case $\uppercase\expandafter{\romannumeral3}$ for odd-even nuclei:
\begin{equation}
\begin{split}
E_{f_{\alpha}}=3B(A-1,Z-1)+B(A-5,Z-3)\\
-2B(A-2,Z-2)-2B(A-2,Z-1),
\end{split}
\end{equation}
\begin{equation}
E=B(A,Z)-B(A-5,Z-3);
\end{equation}

case $\uppercase\expandafter{\romannumeral4}$ for odd-odd nuclei:
\begin{equation}
\begin{split}
E_{f_{\alpha}}=3B(A-2,Z-1)+B(A-6,Z-3)\\
-2B(A-3,Z-2)-2B(A-3,Z-1),
\end{split}
\end{equation}
\begin{equation}
E=B(A,Z)-B(A-6,Z-3).
\end{equation}

\subsection{Semi-empirical formula}
\subsubsection{Universal decay law}
In 2009, using the microscopic mechanism of the charged-particle emission within $\alpha$-like $R$-matrix theory, Qi $et.al$ proposed the universal decay law (UDL) \cite{47,48}. It can be expressed as
\begin{equation}
\log_{10} (T_{1/2})(s)=aZ_{c}Z_{d}\sqrt{\frac{\mathcal{A}}{Q_{c}}}+b\sqrt{\mathcal{A}Z_{c}Z_{d}(A_{c}^{1/3}+A_{d}^{1/3})}+c,          
\end{equation}
where $\mathcal{A}$= $\frac{A_{c}A_{d}}{A_{c}+A_{d}}$ is the reduced mass of the emitted cluster-daughter nucleus system measured in unit of the nucleon mass. $a$=0.4314, $b$=-0.3921, and $c$=-32.7044 are the adjustable parameters.

\subsubsection{Santhosh's semi-empirical model for $\alpha$ decay and cluster radioactivity}
Based on the Geiger-Nuttall (G-N) law, considering the mass asymmetry, Santhosh $et$ $al.$ proposed a formula for estimating the half-lives of $\alpha$ decay and cluster radioactivity in 2008 \cite{49}. It can be given by  
\begin{equation}
\log_{10} (T_{1/2})(s)=aZ_{d}Z_{c}Q_{c}^{-1/2}+b\eta_{A}+c ,     
\end{equation}
where $\eta_{A}$=$\frac{A_{d}-A_{c}}{A}$ is the mass asymmetry. The values of the three adjustable parameters are  $a$=0.727356, $b$=40.3887 and $c$=-85.1625, respectively.

\subsubsection{Unified formula of half-lives for $\alpha$ decay and cluster radioactivity}

Deduced from the WKB barrier penetration probability with some approximation, Ni $et$ $al.$ have proposed a unified formula of half-lives for both $\alpha$ decay and cluster radioactivity \cite{22}. It can be expressed as 
\begin{equation}
\log_{10} (T_{1/2})(s)=a\sqrt{\mathcal{A}}Z_{d}Z_{c}Q_{c}^{-1/2}+b\sqrt{\mathcal{A}}(Z_{d}Z_{c})^{1/2}+c,     
\end{equation}
where $a$=0.38617, $b$=-1.08676, $c_{e-e}$=-21.37195 and $c_{odd-A}$=-20.11223  are the adjustable parameters, respectively.
\section{Results and discussion}
\label{section 3}
Based on Wentzel-Kramers-Brillouin (WKB) theory, considering the cluster preformation probability $P_{c}$, we systematically calculate the cluster radioactivity half-lives of 22 nuclei in the emission of clusters $^{14}$C, $^{15}$N, $^{20}$O, $^{23}$F, $^{24,25,26}$Ne, $^{28,30}$Mg and $^{32,34}$Si from various parent nuclei $^{221}$Fr, $^{221-226}$Ra, $^{223,228,230}$Th, $^{231}$Pa,  $^{232-234}$U, $^{236,238}$Pu and$^{242}$Cm resulting in doubly magic $^{208}$Pb and its neighboring nuclei in this work. The experimental cluster radioactivity half-lives $T^{\rm exp}_{1/2}$ are extracted from Ref.\cite{4} and Ref.\cite{18}. The detailed calculated results 
 are given in Tabel \ref{fig 1}. This table is divided into Part I and Part $\rm \uppercase\expandafter{\romannumeral2}$, which are characterized by $A_{c}<28$ and $A_{c}\ge28$, respectively.  In Tabel \ref{fig 1}, the first to fourth columns represent the decay process, the cluster radioactivity decay energy $Q_{c}$, angular momentum $l$ carried by the emitted cluster and the experimental  cluster radioactivity half-lives in logarithmical form denoted as Decay, $Q_{c}$, $l$ and Exp. The fifth to seventh columns are the calculated results of UDL, Santhosh's semi-empirical model and Ni's formula in logarithmical form denoted as UDL, Santhosh and Ni, respectively. In our work,  $P_{c}$ is calculated through the exponential relationship of $P_{c}$ to $P_{\alpha}$ for $A_{c}$ $<$ 28 with $P_{\alpha}$ obtained by CFM. Nevertheless, since the exponential relationship of $P_{c}$ to $P_{\alpha}$ may break down when $A_{c}\ge$28, $P_{c}$ is calculated through the charge-number dependence of $P_{c}$ on the decay products in DDCM proposed by Ren $et$ $al.$ \cite{9}. The calculated  half-lives results in logarithmical form are given in the eighth column denoted as Cal$^{1}$. For comparison, the theoretical half-life calculations based on WKB theory with $P_{c}$ being calculated through the same charge-number dependence of $P_{c}$ when $A_{c}$ $<$ 28 are also presented in the ninth column denoted as Cal$^{2}$. From this table, it is distinctly to see that the calculated half-lives results using  Ni's formula, Cal$^{1}$, Cal$^{2}$ as well as UDL are basically in agreement with the experimental data. The results in Cal$^{1}$ are better than Cal$^{2}$ when $A_{c}$ $<$ 28. 

For the sake of intuitively comparing these results, we plot the differences between the experimental cluster radioactivity half-lives  and the calculated ones by using UDL, Santhosh's model, Ni's formula, Cal$^{1}$ and Cal$^{2}$ in logarithmical form in Fig. \ref{fig 1}. From this figure, it is distinctly to see that the differences between the experimental data and the calculated results in Cal$^{1}$ and Ni's formula are within $ \pm2$ on the whole, showing the calculated cluster radioactivity half-lives in Cal$^{1}$ as well as Ni's formula can reproduce the experimental data well.  By contrast, in the case of $^{223}$Ra$\to$$^{209}$Pb+$^{14}$C, there is a discrepancy of 2.482 in UDL. For $^{223}$Ac$\to$$^{209}$Bi+$^{14}$C, there is a discrepancy of 2.325 in Cal$^{2}$. While the values of the discrepancies of 7 nuclei out of the whole 22 nuclei are out of the scale of $\pm$ 2 within Santhosh's model. Especially, as for $^{231}$Pa$\to$$^{208}$Pb+$^{23}$F and $^{242}$Cm $\to$$^{208}$Pb+$^{34}$Si, the discrepancies can even achieve at 4.31 and 5.71, respectively. Furthermore, from the overall trend, the results in Cal$^{1}$ are more converging on the neighboring zero line area, relatively similar to the distribution of UDL and Ni's formula. Nevertheless, the distribution for the results of Santhosh's model is slightly scattered. For further investigating the consistencies between the cluster radioactivity half-lives experimental data and the calculated ones obtained by UDL, Santhosh's model, Ni's formula, Cal$^{1}$ and Cal$^{2}$, the standard deviation $\sigma$ is used, which is defined by 
\begin{equation}
\sigma=[\sum_{i}^{n}(\log_{10}T^{\rm exp}_{1/2_{i}}-\log_{10}T^{\rm cal}_{1/2_{i}})^2/n]^{1/2},  	     
\end{equation}
where $\log_{10}T^{\rm exp}_{1/2_{i}}$ and $\log_{10}T^{\rm cal}_{1/2_{i}}$ denote the logarithmical form of the experimental  cluster radioactivity half-lives and calculated ones for the $i$-th nucleus, respectively. The calculated results of $\sigma$ using UDL,  Santhosh's model, Ni's formula, Cal$^{1}$ and Cal$^{2}$ are presented in Table \ref{table 2}. From this table, as $A_{c}$ $<$ 28, the $\sigma$ value between the experimental data and the results obtained in CFM is 1.035, smaller comparing to the results from Cal$^{2}$, UDL and  Santhosh's model which are 1.256, 1.244 and 1.880, respectively, larger than the results from Ni with 0.365. As $A_{c}$ $\ge$ 28, the $\sigma$ value of Cal$^{1}$ and Ni's formula are 0.423 and 0.425, smaller than the $\sigma$ value from UDL and Santhosh's model, which are 0.569 and 3.594, respectively. For the total nuclei, it is clearly to see that $\sigma$ between the experimental data and the results calculated in Cal$^{1}$ is 0.910, achieving better than those resulting from UDL and Santhosh's model, which are 1.102 and 2.469, respectively, more closer to Ni's formula with 0.382 which is the least. Moreover, the calculated results in Cal$^{1}$ are more better than the ones based on different types of Proximity potentials from Ref.\cite{71},  where the standard deviation $\sigma$ between experimental and calculated half-lives from 1.373 up to 7.951. This 
indicates that the our calculated cluster radioactivity half-lives results  can reproduce the experimental data well. However, as for $^{231}$Pa$\to$$^{208}$Pb+$^{23}$F with a discrepancy of 1.962 and  $^{233}$U$\to$  $^{209}$Pb+$^{24}$Ne with a discrepancy of 1.825, the experimental data can not be properly reproduced. This reason may be accounted for the imperfection of early detection technologies and radioactive beam equipments.

 	\begin{center}
 	\includegraphics[width=7.9cm]{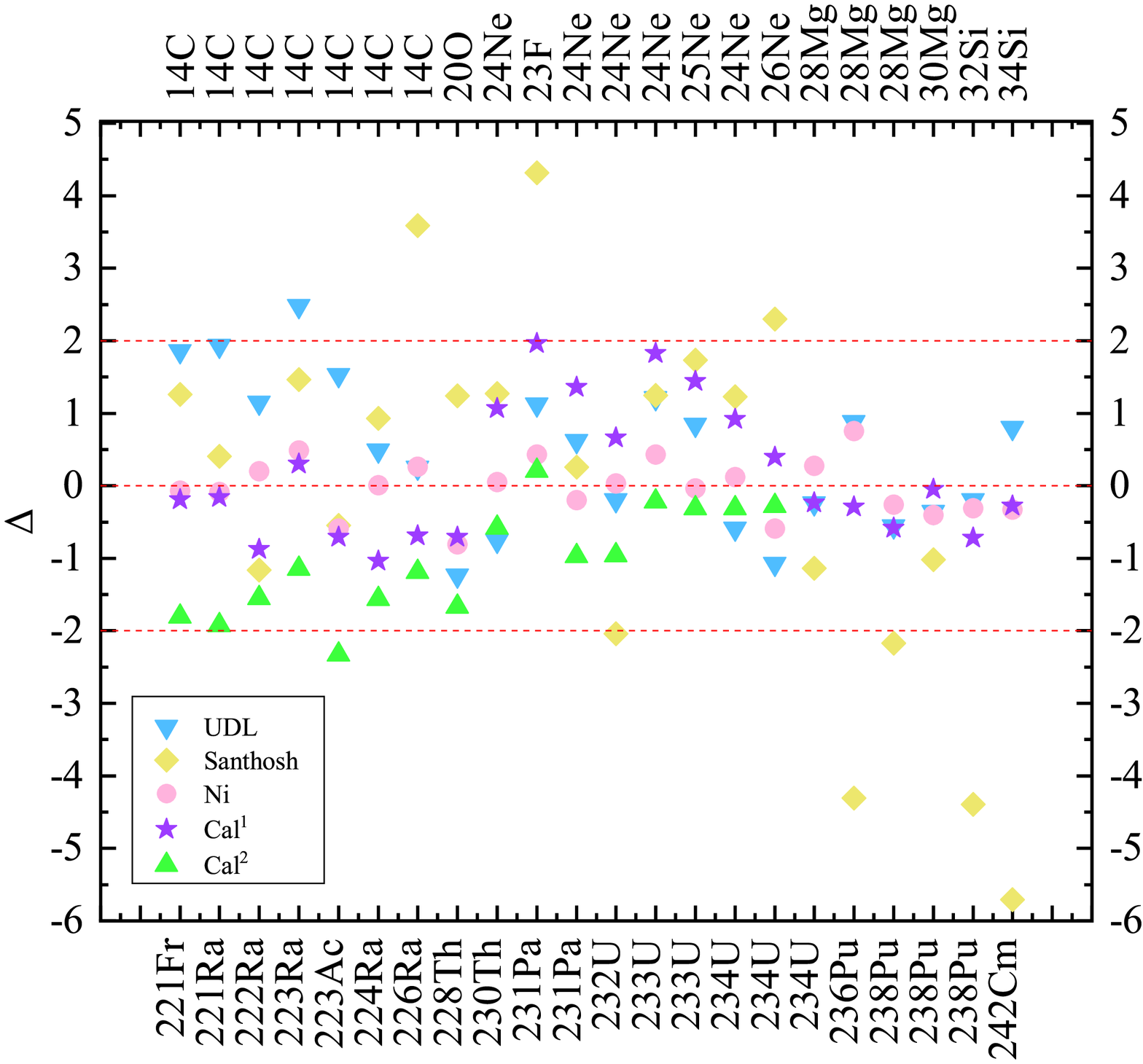}
 	\figcaption{\label{figure1}Comparison for the discrepancy in logarithmical form between the experimental cluster radioactivity half-lives and calculated ones obtained in UDL, Santhosh's model, Ni's formula and our model.}
 	\label {fig 1}
 \end{center}

Encouraged by the good agreement between the experimental cluster radioactivity half-lives and calculated ones within our model. In the following, we extend this model to predict the half-lives for the  possible cluster   radioactive candidates. The detailed calculated results are given in Table \ref{table 3}. In this table, the indications for the first to seventh columns are similar to Table \ref{table 1}. From Table \ref{table 3}, it is obviously to see that the predicted results in Cal$^{1}$ are more closer to those predicted by using UDL and Ni's formula except for the ones predicted by using Santhosh's model. Note that most of the predicted results are of the same order of magnitude. For instance, in the case of $^{227}$Th$\to$ $^{209}$Pb+$^{18}$O, the predicted cluster radioactivity half-lives using UDL, Santhosh's model, Ni's formula and Cal$^{1}$ are 21.00, 20.59, 21.69 and 21.38, respectively. It implies that our predictions are reliable. This work may provide theoretical direction for future experiment.

Besides, the correlations between the cluster  radioactivity half-lives in logarithmical form log$_{10}$$T_{1/2}$, penetration probability in logarithmical form  log$(P)$ versus daughter neutron number are plotted in Figs. (\ref{fig 2} to \ref{fig 5}). Fig. \ref{fig 2} and  Fig. \ref{fig 4}. represent the cluster radioactivity half-lives in logarithmical form versus the daughter neutron number for the $^{14}$C emission from Ra isotopes and $^{24}$Ne emission from Ra isotopes, respectively. Fig. \ref{fig 3} and Fig. \ref{fig 5} denote the cluster radioactivity penetration probability in logarithmical form versus the daughter neutron number for the $^{14}$C emission from Ra isotopes and $^{24}$Ne emission from Ra isotopes, respectively. From Fig. \ref{fig 2} and  Fig. \ref{fig 4}, it is clearly to see with the increasing of the daughter neutron number, log$_{10}$$T_{1/2}$ begin to decrease almost linearly. When it arrives at the magic neutron number 126, the value of the log$_{10}$$T_{1/2}$ reaches the minimum, then increase linearly again. This is exactly opposite for the tendency for the log$(P)$ depicted in Fig. \ref{fig 3} and Fig. \ref{fig 5}. Nevertheless, when the daughter neutron number at doubly magic daughter nucleus $^{208}$Pb, the log$_{10}$$T_{1/2}$ and log$(P)$ both take minimum and maximum values. Consequently, it confirms that neutron shell closure plays a more crucial role than proton shell closure in cluster radioactivity, revealing the neutrons paring effect is more influential than protons in this decay mode process \cite{5}.

Furthermore, recent works have shown that the dynamical deformations of the shapes of both the cluster and daughter nucleus are also important in cluster radioactivity \cite{63,64,65,67,72,73}. The barrier penetration can be changed due to the shape deformations of both the cluster and the residue nucleus at the cluster emission process \cite{63}. The deformations can be reflected in orientation angle dependent nuclear radius resulting in the changes of barrier heights \cite{65,72}, which could exert the influence in the penetrability, moreover, causing alteration to the cluster radioactivity half-lives \cite{64,67,73}. Therefore, it is meaningful to consider dynamical shape deformations of both the cluster and daughter nucleus to study this decay mode. And we will investigate the significance of the dynamical deformations of the shapes of both the cluster and daughter nucleus in cluster radioactivity in more details in our future work.
\end{multicols}

\begin{center}
	\tabcaption{Comparisons between the experimental cluster radioactivity half-lives (in seconds) and the calculated ones using UDL, Santhosh's model, Ni's formula and our model in logarithmical form.}
	\label {table 1}
	\footnotesize
	\begin{tabular}{ccccccccc}
		\hline \hline
		\multicolumn{2}{c}{\multirow{3}{*}{}}&\multicolumn{7}{c}{$\log_{10}T_{1/2}$ (s)}\\
		\cline{4-9} 
		{Decay} & $Q_{c}$ (MeV)& {$l$} &{Exp} & {UDL}&{Santhosh}&Ni&Cal$^{1}$ & Cal$^{2}$\\ \hline 
		\noalign{\global\arrayrulewidth1.2pt}\noalign{\global\arrayrulewidth0.4pt} \multicolumn{9}{c}{\textbf{}}\\
			\noalign{\global\arrayrulewidth1pt}\noalign{\global\arrayrulewidth0.4pt} \multicolumn{9}{c}{\textbf{ Part I: $A_{c}<28$}}\\
		$^{221}$Fr$\to$ $^{207}$Tl+$^{14}$C	&$31.2911$	 &	$3$  &	$14.56$ 	&$	12.70     $&$13.30     $&$14.63$&$	14.76 	$&$16.37
			$
		\\  
		$^{221}$Ra$\to$$^{207}$Pb+$^{14}$C      &$32.3961$ &    $3$&  $13.39$  &$	11.46	$&$	12.98	$&$13.48$&$	13.55	$&$15.31 
			$\\
		$^{222}$Ra$\to$$^{208}$Pb+$^{14}$C      &$33.0486$ &    $0$&  $11.22$  &$	10.07	$&$	12.38	$&$11.02$&$	12.10	$&$12.77$\\
		$^{223}$Ra$\to$$^{209}$Pb+$^{14}$C	&     $31.8279$   &  $2$  &$15.05 	$&$	12.57	$&$13.59	$&$14.56$
		&$	14.75 	$&$16.20$\\
	
		$^{224}$Ra$\to$$^{210}$Pb+$^{14}$C  &$30.5343$ 	&     $0$   &  $15.87$ &$	15.38 	$&$14.94	$&$15.86$ &$	16.91	$&$17.43$ \\
		$^{226}$Ra$\to$ $^{212}$Pb+$^{14}$C      &$28.1966$ 	&     $0$  &  $21.20$  &$	20.95 	$&$17.62	$&$20.94$&$	21.89 	$&$22.39$\\ 
			$^{223}$Ac$\to$$^{209}$Bi+$^{14}$C    &$33.0636$ 	&     $2$   &  $12.60$  &$	11.07 	$&$13.15	$&$13.19$&$	13.31 	$&$14.93$\\   
		$^{228}$Th$\to$$^{208}$Pb+$^{20}$O  &$44.7233$ 	&     $0$   &  $20.73$  &$	21.97	$&$19.49	$&$21.54$&$	21.44	$&$22.39$\\ 
			$^{231}$Pa$\to$$^{208}$Pb+$^{23}$F          &$51.8828$	&     $1$     &   $26.02$  &$24.90	$&$21.71	$&$25.59$&$	24.06	$&$22.39$\\ 
		$^{230}$Th$\to$$^{206}$Hg+$^{24}$Ne   &$57.7599$ 	&   $0$   &   $24.63$  &$	25.39	$&$23.36	$&$24.58$&$	23.57	$&$25.82$\\ 
	
		$^{231}$Pa$\to$$^{207}$Tl+$^{24}$Ne         &$60.4099$	&     $1$    &   $22.89$  &$	22.27	$&$22.64	$&$23.09$&$	21.54	$&$25.21$\\
		$^{232}$U$\to$ $^{208}$Pb+$^{24}$Ne &$62.3095$ 	& $0$     &   $20.39$ &$	20.59	$&$22.43    $&$20.36$ &$	19.73 	$&$23.86$\\  
		$^{233}$U$\to$ $^{209}$Pb+$^{24}$Ne   &$60.4853$ 	&     $2$     &   $24.84$ &$	23.63	$&$23.60	$&$24.41$ &$	23.02	$&$21.35$ \\ 
		 
		$^{234}$U$\to$$^{210}$Pb+$^{24}$Ne    &$58.8250$ 	&     $0$    &   $25.93$  &$26.52	$&$24.71	$&$25.81$&$25.01	$&$25.06$\\ 
		$^{233}$U$\to$ $^{208}$Pb+$^{25}$Ne       &$60.7036$	&     $2$     &   $24.84$ &$	24.00	$&$23.11	$&$24.88$ &$	23.40	$&$25.15$\\ 
		$^{234}$U$\to$$^{208}$Pb+$^{26}$Ne       &$59.4125$ 	&     $      0$   &   $25.93$  &$27.01$&$23.63$&$26.52$ &$
		25.53$&$26.21$\\  
			
				\noalign{\global\arrayrulewidth1pt}\noalign{\global\arrayrulewidth0.4pt} \multicolumn{9}{c}{\textbf{ Part I: $A_{c}\ge28$}}\\
		$^{234}$U$\to$$^{206}$Hg+$^{28}$Mg    &$74.1108$ 	&     $0$    &   $25.53$  &$	25.77	$&$26.67	$&$25.25$&$25.76$&$-$\\ 
		$^{236}$Pu$\to$$^{208}$Pb+$^{28}$Mg&$79.6700$&$0$&$21.52$&$20.64$&$25.83$&$20.76$ &$21.81$&$	-	$\\
		$^{238}$Pu$\to$$^{210}$Pb+$^{28}$Mg&$75.9114$&$0$&$25.70$&$26.26$&$27.87$&$25.96$ &26.28&$	-	$
		\\
		$^{238}$Pu$\to$$^{208}$Pb+$^{30}$Mg&$76.7930$&$0$&$25.70$&$26.06$&$26.72$&$26.10 
			$&$25.75$&$-$
		\\
		$^{238}$Pu$\to$$^{206}$Hg+$^{32}$Si&$96.1867$&$0$&$25.28$&$25.48$&$29.68$&$25.59$&$26.00$&$-$
		\\
		$^{242}$Cm $\to$$^{208}$Pb+$^{34}$Si&$96.5440$&$0$&$23.15$&$22.35$&$28.86$&$23.48$&$23.43$&$-$
		\\
		\hline \hline
	\end{tabular}
\end{center}

\begin{center}
	\tabcaption{Standard deviation $\sigma$ between the experimental data and the calculated ones using different theoretical models and/or formulas for cluster radioactivity.}
	\label {table 2}
	\footnotesize
	\begin{tabular}{cccccc}
		\hline
		{Model} & {UDL} &{Santhosh}&Ni&Cal$^{1}$& Cal$^{2}$ \\ \hline 
		
		$\sigma(A_{c}<28)$ 
		&1.244
		&1.880
		&0.365
		&1.035&1.256
		 \\
		$\sigma(A_{c}\ge28)$ 
		&0.569
		&3.594
		&0.425
		&0.423
		&$-$
		 \\
		$\sigma$  &$1.102$&$2.469$&$0.382$&$0.910$&$-$\\
		
		\hline 
	\end{tabular}
\end{center}
\begin{multicols}{2}
	
	\begin{center}
 		\includegraphics[width=7.3cm]{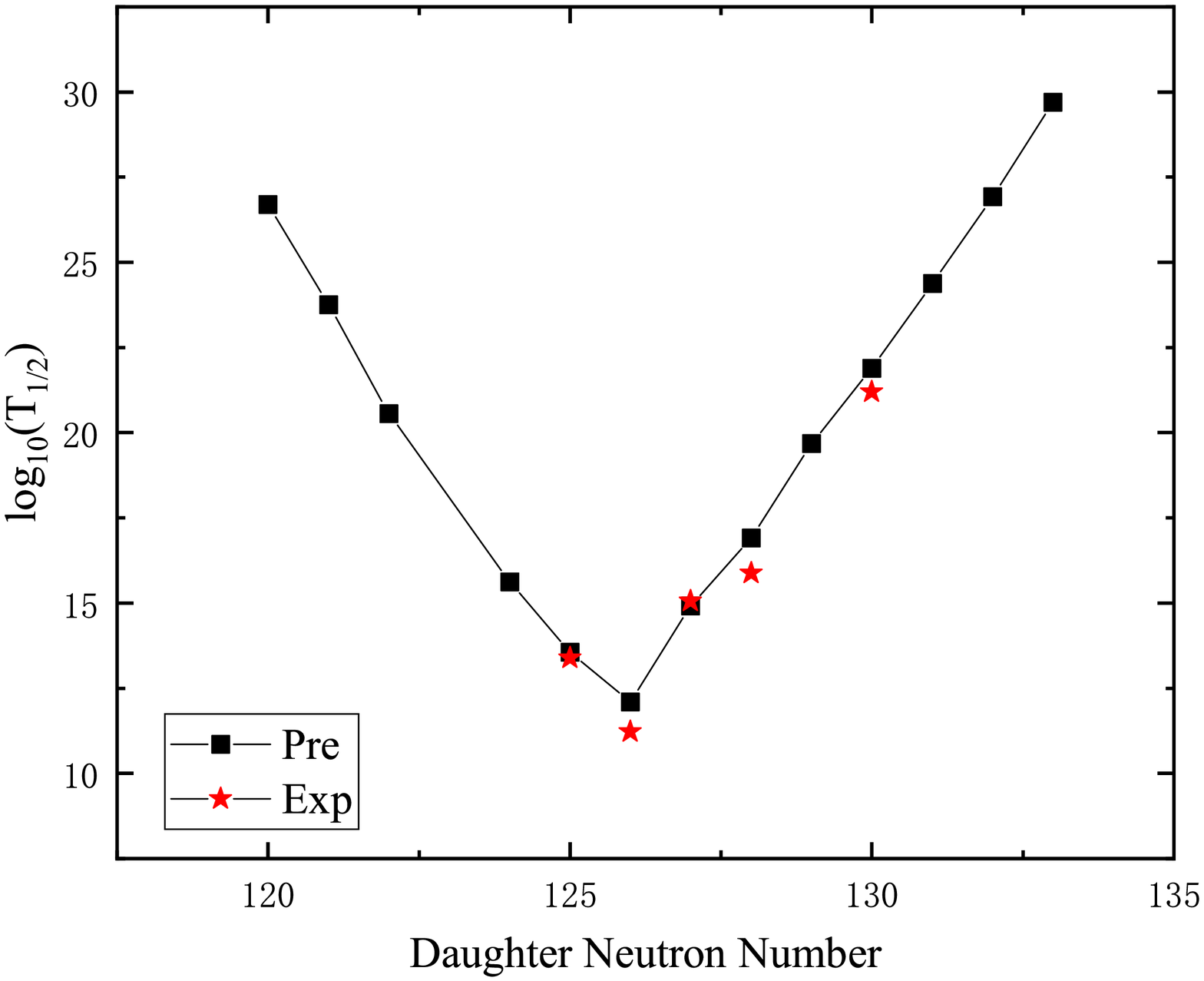}
		\figcaption{\label{figure5}(color online) Plot of the computed log$_{10}$($T_{1/2}$) values versus neutron number of daughter for the emission of cluster $^{14}$C from Ra isotopes. The black squares and red pentagrams represent the predicted and experimental half-lives, respectively.}
		\label {fig 2}	
	\end{center}

	\begin{center}
		\includegraphics[width=7.3cm]{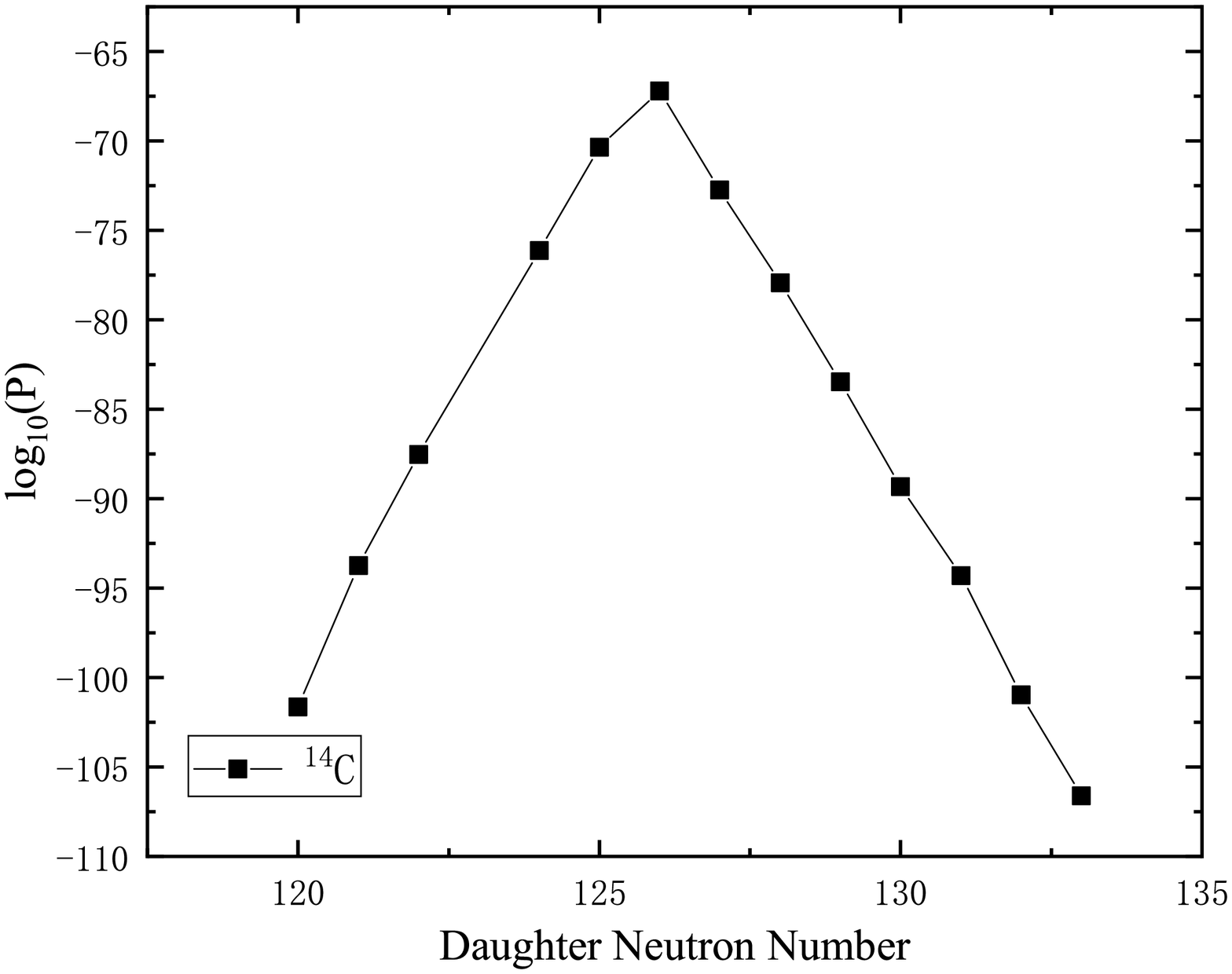}
		\figcaption{\label{figure5} Plot of the computed log$_{10}$$(P)$ values versus neutron number of daughter for the emission of cluster $^{14}$C from Ra isotopes. }
		\label {fig 3}	
	\end{center}

 \end{multicols}

\begin{center}
	\tabcaption{Predicted half-lives for possible cluster radioactive nuclei.}
	\label {table 3}
	\footnotesize
	\begin{tabular}{cccccccc}
		\hline \hline
		\multicolumn{2}{c}{\multirow{3}{*}{}}&\multicolumn{5}{c}{$\log_{10}T_{1/2}$ (s)}\\
		\cline{4-8} 
		{Decay} & $Q_{c}$ (MeV) & \emph{l}  &{Exp}  & {UDL}&{Santhosh}&{Ni}&{Cal$^{1}$}\\ \hline 
		\noalign{\global\arrayrulewidth1.2pt}\noalign{\global\arrayrulewidth0.4pt} \multicolumn{8}{c}{\textbf{}}\\
		$^{219}$Rn$\to$ $^{205}$Hg+$^{14}$C	&$28.0974 
		$ & $3$&	- 	   	&$	19.09 	$&$	15.93   $ &$20.29$ &	$20.41 $\\  
		$^{220}$Rn$\to$ $^{206}$Hg+$^{14}$C &$28.5380 
		$ & $0 $ &- 	  &	$17.95 $ 	&$15.44 
		 $ &$18.05$&	$18.98 $\\  
		$^{221}$Fr$\to$ $^{206}$Hg+$^{15}$N &$34.1215 
		$ & $3 $ &-	 	  &	$21.55  $ 	&$19.47 
		 $ &$22.16$&$22.06  $\\ 
		$^{223}$Ra$\to$ $^{205}$Hg+$^{18}$O &$40.3040 
		$ & $1$&	- 	  	&$	26.44 	$&$	22.03 
		   $ &$26.41$&	$25.88   $ \\
		$^{225}$Ra$\to$ $^{211}$Pb+$^{14}$C	&$29.4661
		$ & $4$&	- 	  	&$	17.84 	$&$	16.13 
		   $ &$19.37$ &	$19.68  $\\ 
		$^{225}$Ra$\to$ $^{205}$Hg+$^{20}$O	&$40.4847 
		
		$ & $1$&	- 	  	&$	28.27  	$&$	21.21 
		 $ &$28.34$ &	$27.59  $ \\ $^{226}$Ra$\to$ $^{206}$Hg+$^{20}$O	&$40.8173 
		
		$ & $0$&	- 	   	&$	27.46 	$&$	20.94 
		  $ &$26.41$&	$26.34  $ \\ $^{223}$Ac$\to$ $^{208}$Pb+$^{15}$N	&$39.4720 
		$ & $3$&	$>14.76$ 	   	&$	12.93  	$&$	16.25 
		 $&$14.47$ &	$14.53  $\\ $^{227}$Ac$\to$ $^{207}$Tl+$^{20}$O	&$43.0872  
		$ & $1$&	- 	  	&$	23.95 
		$&$	19.91 
		$ &$24.55$   &	$23.78  $ \\ $^{229}$Ac$\to$ $^{206}$Hg+$^{23}$F	&$48.3458 
		
		$ & $2$&	- 	   	&$	28.93 	$&$	22.43 
		  $&$29.14$&	$27.69 $ \\ $^{226}$Th$\to$ $^{208}$Pb+$^{18}$O	&$45.7293 
		
		$ & $0$&	- 	  	&$	18.14 	$&$	19.35 
		 $&$17.81$ &	$18.22  $ \\ $^{226}$Th$\to$ $^{212}$Po+$^{14}$C	&$30.5475  
		$ & $0$&	$>16.76$ 	 	&$17.55 	$&$	16.55 
		  $ &$17.83$  &	$18.78  $ \\ $^{227}$Th$\to$ $^{209}$Pb+$^{18}$O	&$44.2021 
		
		$ & $4$&	$>15.30 $	   	&$	21.00 	$&$	20.59 
		  $ &$21.69$&	$21.38  $\\ $^{228}$Th$\to$ $^{206}$Hg+$^{22}$Ne	&$55.7416 
		
		$ & $0$&	- 	   	&$	27.48  	$&$	25.37 
		  $&$26.07$&	$25.34   $ \\ $^{229}$Th$\to$ $^{209}$Pb+$^{20}$O	&$43.4038 
		
		$ & $2$&	- 	   	&$	24.64 	$&$	20.60

		  $ &$25.25$&	$24.34  $
		  \\ $^{229}$Th$\to$ $^{205}$Hg+$^{24}$Ne	&$57.8251 
		
		$ & $3$&	- 	   	&$25.34 	$&$	23.28 
		 $ &$25.72$&	$24.30  $\\ $^{231}$Th$\to$ $^{207}$Hg+$^{24}$Ne	&$56.2544 
		
		$ & $2$&	- 	  	&$	28.12 
		$&$	24.42 
		
		$ &$28.36$ &	$26.54  $ \\ $^{231}$Th$\to$ $^{206}$Hg+$^{25}$Ne	&$56.7977 
		
		$ & $2$&	- 	  	&$	27.92 
		$&$	23.69 
		
		$  &$28.30$&	$26.42 
		$ \\ $^{232}$Th$\to$ $^{208}$Hg+$^{24}$Ne	&$54.6683 
		$ & $0$&	$>29.20$ 	  &$	31.12 $&$	25.57 
		   $ &$29.86$&	$28.81 
		   $ 	 \\ $^{232}$Th$\to$ $^{206}$Hg+$^{26}$Ne	&$55.9116 
		
		$ & $0$&	$>29.20 $	  	&$	30.37 	$&$	23.99 
		   $ &$29.45$ &	$28.25  $ \\ $^{227}$Pa$\to$ $^{209}$Bi+$^{18}$O	&$45.8713 
		
		$ & $2$&	- 	  	&$19.16 
		$&$	20.13 
		 $ &$20.01$  &	$19.48 $\\ $^{229}$Pa$\to$ $^{207}$Tl+$^{22}$Ne	&$58.9558 
		
		$ & $2$&	- 	  	&$23.31 	$&$	24.20 
		   $ &$23.63$ &	$22.31  $\\ $^{230}$U$\to$ $^{208}$Pb+$^{22}$Ne	&$61.3883 
		
		$ & $0$&	$>18.20$ 	&$	20.73 	$&$	23.62 
		  $  &$20.09$ &	$19.49  $\\ $^{230}$U$\to$ $^{206}$Pb+$^{24}$Ne	&$61.3521 
		
		$ & $0$&$>18.20$   	&$22.34 	$&$	22.94 
		  $  &$21.78$&	$20.88  $ \\ $^{232}$U$\to$ $^{204}$Hg+$^{28}$Mg	&$74.3195 
		
		$ & $0$&	$>22.26$	 	&$	25.59 	$&$	26.47 
		  $ &$24.93$ &	$23.45  $ \\ $^{233}$U$\to$ $^{205}$Hg+$^{28}$Mg	&$74.2271 
		
		$ & $3$&$>27.59$   	&$	25.66  	$&$	26.57 
		  $ &$26.33$&	$24.42  $  \\ $^{235}$U$\to$ $^{211}$Pb+$^{24}$Ne	&$57.3635 
		
		$ & $1$& $>27.65$    	&$	29.16 	$&$	25.73 
		  $& $29.51$&	$27.91  $\\ $^{235}$U$\to$ $^{210}$Pb+$^{25}$Ne	&$57.6832 
		
		$ & $3$&	$>27.65$	 	&$	29.41 	$&$	25.16 
		 $ &$29.86$  &	$28.25  $ \\ $^{235}$U$\to$ $^{207}$Hg+$^{28}$Mg	&$72.4257 
		
		$ & $1$&$>28.45$  	&$	28.45 	$&$	27.65 
		  $ &$29.00$ &	$26.98  $\\ $^{235}$U$\to$ $^{206}$Hg+$^{29}$Mg	&$72.4772 
		
		$ & $3$&$>28.45$  	&$	29.03 	$&$27.28 
		 $ &$29.67$ &	$29.44  $ \\ $^{236}$U$\to$ $^{212}$Pb+$^{24}$Ne&$55.9451 
		
		$ & $0$&$>26.27$   	&$	31.83 	$&$	26.75 
		$ &$30.71$ &	$29.38 $\\ $^{236}$U$\to$ $^{210}$Pb+$^{26}$Ne&$56.6920 
		
		$ & $0$&$>26.27$  	&$32.10 	$&$	25.54 
		  $ &$31.23$  &	$29.70  $\\ $^{236}$U$\to$ $^{208}$Hg+$^{28}$Mg	&$70.7345 
		
		$ & $0$&	$>26.27$	  	&$	31.25 	$&$	28.67 
		 $ &$30.33$&	$28.36  $ \\ $^{236}$U$\to$ $^{206}$Hg+$^{30}$Mg	&$72.2719 
		
		$ & $0$&$>26.27$  	&$29.94  	$&$	27.09 
		   $ &$29.47$&	$28.74 $ \\ $^{238}$U$\to$ $^{208}$Hg+$^{30}$Mg	&$69.4591 $ & $0$&	- 	  	&$	34.79 
		$&$	28.83 
		
		$  &$33.98$&	$32.67  $ \\ $^{231}$Np$\to$ $^{209}$Bi+$^{22}$Ne	&$61.9033 
		
		$ & $3$&	- 	  &$	21.37 
		$&$	24.26 
		
		$ &$21.95$ &	$20.47 
		$ 	\\ $^{233}$Np$\to$ $^{209}$Bi+$^{24}$Ne	&$62.1600 
		
		$ & $3$&	- 	  	&$	22.37 	$&$	23.48 
		  $ &$23.24$&	$21.75 
		  $ \\ $^{235}$Np$\to$ $^{207}$Tl+$^{28}$Mg	&$77.0969 
		
		$ & $2$&	- 	  	&$	22.82 
		$&$	26.12 
		 
		$ &$23.92$&	$22.27 
		$ \\ $^{237}$Np$\to$ $^{207}$Tl+$^{30}$Mg	&$74.7869 
		
		$ & $2$&	$>27.57$	 	&$	27.54 
		$&$	26.75 
		
		$  &$28.63$  &	$28.06 
		$ \\ $^{237}$Pu$\to$ $^{209}$Pb+$^{28}$Mg	&$77.7263 
		
		$ & $1$&	- 	 	&$23.49 $&$	26.86 
		  $ &$24.66$ &	$23.23 
		  $  \\ $^{237}$Pu$\to$ $^{208}$Pb+$^{29}$Mg	&$77.4527 
		
		$ & $3$&	- 	 &$	24.51	$&$	26.67 
		   $&$25.73$ &	$25.94 
		   $ 	 \\ $^{237}$Pu$\to$ $^{205}$Hg+$^{32}$Si&$91.4574 
		
		$ & $4$&	- 	 &$	25.17  
		$&$	29.50 
		 
		$  &$26.48$ &	$27.04 
		$ 	\\ $^{239}$Pu$\to$ $^{209}$Pb+$^{30}$Mg	&$75.0841 
		
		$ & $4$&	- 	  	&$	28.78 
		$&$	27.68 
		 
		$&$29.89$&	$29.20
		$ \\ $^{239}$Pu$\to$ $^{205}$Hg+$^{34}$Si	&$90.8678 
		
		$ & $1$&	- 	  &$	26.83 
		$ &$29.19 
		   $ &$28.50$&	$27.90
		   $\\ $^{237}$Am$\to$ $^{209}$Bi+$^{28}$Mg	&$79.8484 
		
		$ & $2$&	- 	   	&$	22.06 $&$26.76 
		    $ &$23.35$ &	$21.75   $\\ $^{239}$Am$\to$ $^{207}$Tl+$^{32}$Si&$94.5021 $ & $3$&	- 	  	&$	22.65  
		$&$	29.26 
		 $  &$24.38$&	$25.11  $ \\ $^{241}$Am$\to$ $^{207}$Tl+$^{34}$Si	&$93.9599 
		
		$ & $3$&	$>24.41$ 	   	&$	24.13 	$&$	28.92 
		$ &$26.26$&	$25.91  $\\ $^{240}$Cm$\to$ $^{208}$Pb+$^{32}$Si&$97.5504 
		
		$ & $0$&	- 	  	&$	20.31  	$&$	29.00 
		 $ &$21.09$&	$22.19  $ \\ $^{241}$Cm$\to$ $^{209}$Pb+$^{32}$Si&$95.3940 
		
		$ & $4$&	- 	   	&$	23.19 	$&$	29.99 
		
		$  &$25.02$&	$25.65   $\\ $^{243}$Cm$\to$ $^{209}$Pb+$^{34}$Si	&$94.7881 
		
		$ & $2$&	- 	  	&$	24.77 
		$&$	29.69 
		
		$  &$27.00$&	$26.47 
		$ \\ $^{244}$Cm$\to$ $^{210}$Pb+$^{34}$Si	&$93.1718 
		
		$ & $0$&	- 	  &$	27.06 
		$&$	30.48 
		
		$&$27.89$&	$27.03
		$ 	\\ 
		\hline \hline
		\end{tabular}
		\end{center}

\begin{multicols}{2}

\begin{center}
	\includegraphics[width=7.3cm]{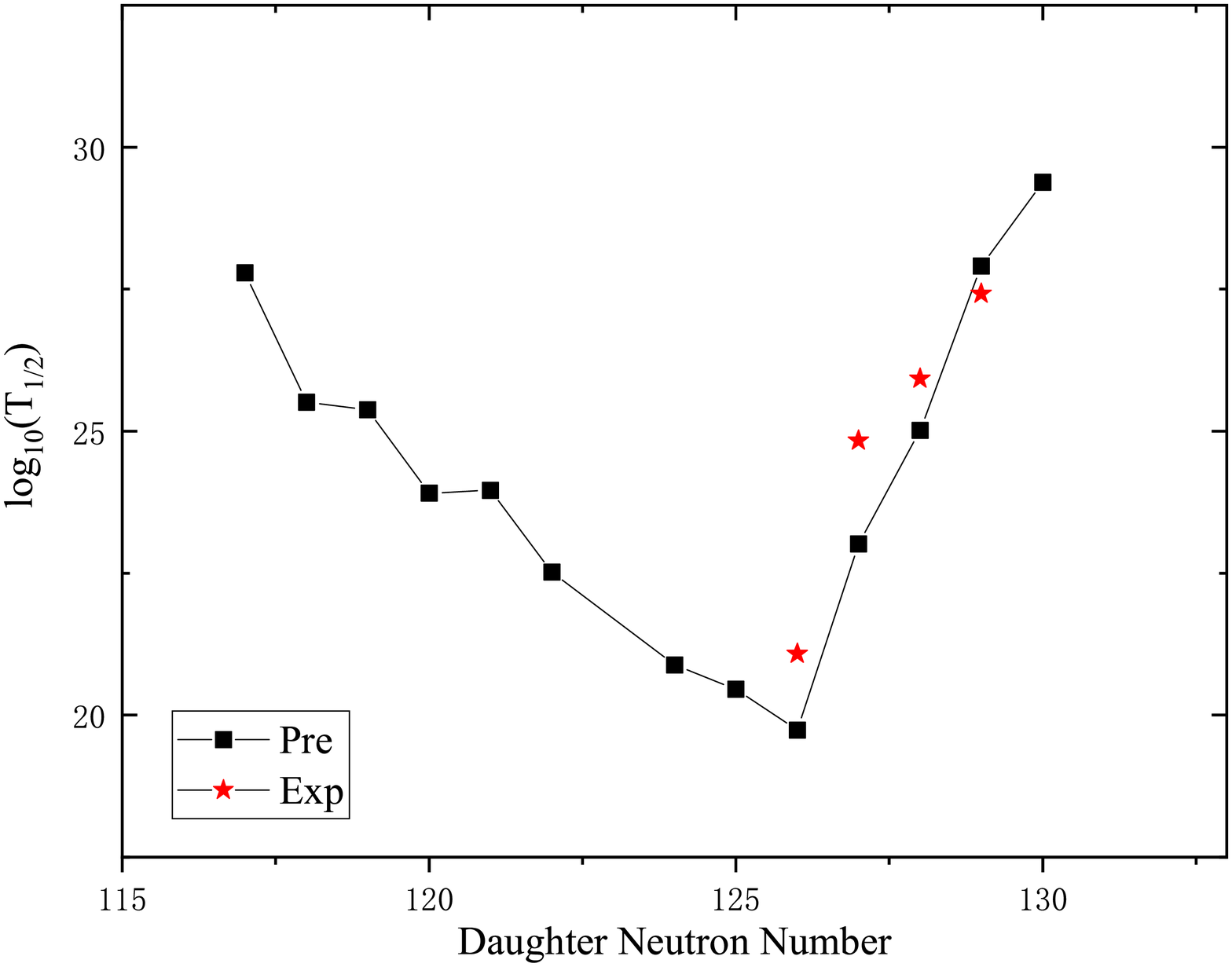}
	\figcaption{\label{figure5} (color online) Plot of the computed log$_{10}$($T_{1/2}$) values versus neutron number of daughter for the emission of cluster $^{24}$Ne from Ra isotopes. The black squares and red pentagrams represent the predicted and experimental half-lives, respectively.}
	\label {fig 4}	
\end{center}

\begin{center}
	\includegraphics[width=7.3cm]{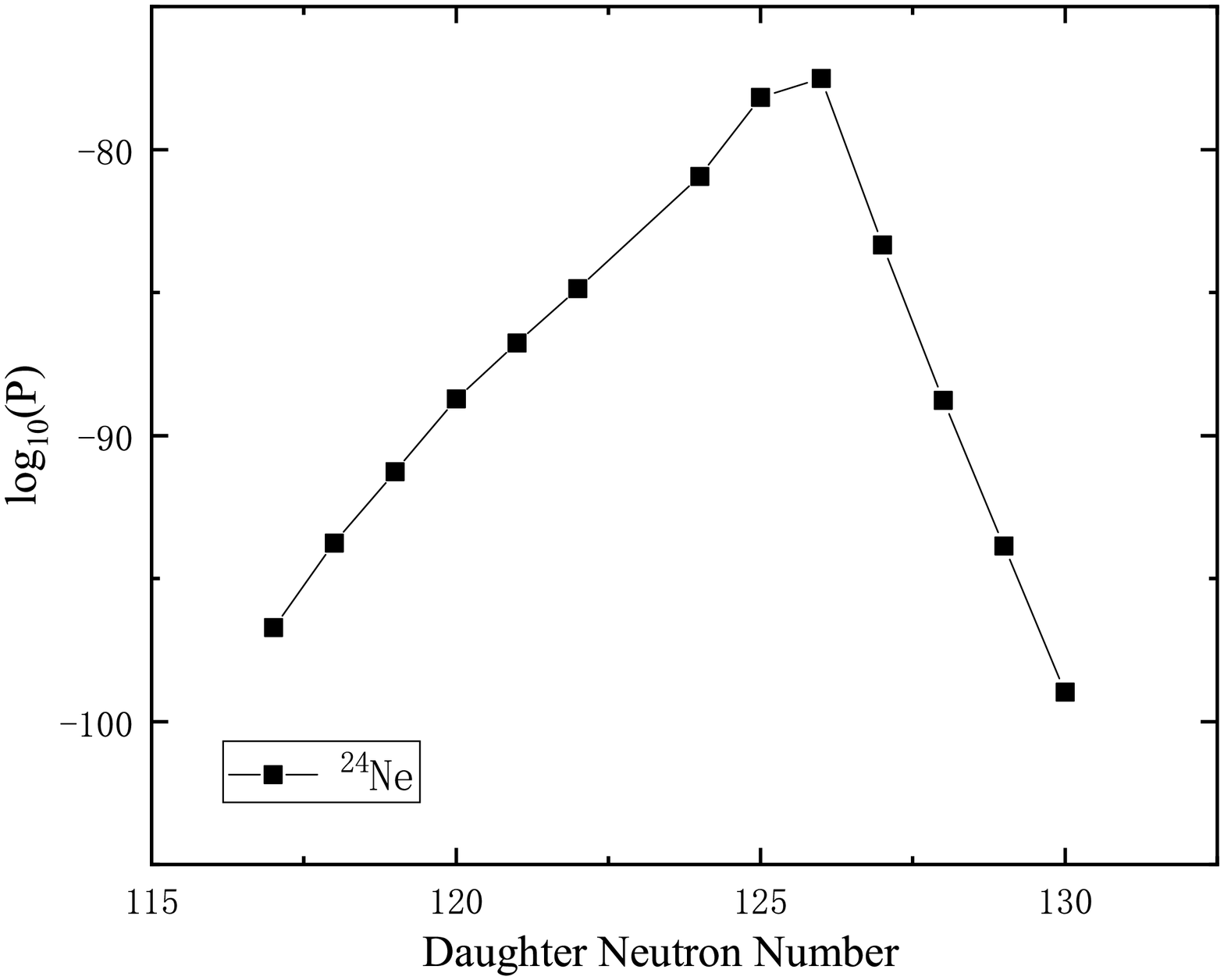}
	\figcaption{\label{figure5} Plot of the computed log$_{10}$$(P)$ values versus neutron number of daughter for the emission of cluster $^{24}$Ne from Ra isotopes. }
	\label {fig 5}	
\end{center}
 \end{multicols}

\begin{multicols}{2}
\section{Summary}
\label{section 4}
In summary, based on the WKB theory, considering the cluster radioactivity preformation probability $P_{c}$ and using a Woods-Saxon type nuclear potential, we systematically study the half-lives of 22 nuclei experimentally observed. The preformation factor $P_{c}$ is obtained within CFM applied with the exponential relationship of $P_{\alpha}$ when $A_{c}$ $<$ 28 and an effective empirical formula when $A_{c}$ $\ge$ 28. The calculated cluster radioactivity half-lives are compared with UDL, Santhosh's semi-empirical model for both $\alpha$ decay and cluster radioactivity and Ni's unified formula for both $\alpha$ decay and cluster radioactivity. The results are consistent with the experimental data.  Moreover, we extend this model to predict the cluster radioactivity half-lives for the possible candidates. Finally, we confirm the neutron magicity at daughter neutron number 126 and the neutrons paring effect is more influential than protons in cluster radioactivity. 
\end{multicols}
\vspace{-1mm}
\centerline{\rule{80mm}{0.1pt}}
\vspace{2mm}

\begin{multicols}{2}

\end{multicols}

\clearpage
\end{CJK*}
\end{document}